\documentclass[12pt, a4paper]{article}
\usepackage{amsfonts}
\usepackage{amsmath}
\usepackage{natbib}
\usepackage{graphicx}
\usepackage{float}
\usepackage{PRIMEarxiv}
\usepackage{algorithm}
\usepackage{algpseudocode}

\usepackage[utf8]{inputenc} 
\usepackage[T1]{fontenc}    
\usepackage{hyperref}       
\usepackage{url}            
\usepackage{booktabs}       
\usepackage{nicefrac}       
\usepackage{microtype}      
\usepackage{fancyhdr}       
\usepackage{graphicx}       
\graphicspath{{media/}}     
\usepackage{setspace}
\usepackage{xcolor}
\usepackage{listings}

\newcommand{\vh}{\mathbf{h}}
\newcommand{\vH}{\mathbf{H}}

\newcommand{\vO}{\mathbf{O}}

\newcommand{\vU}{\mathbf{U}}

\newcommand{\vV}{\mathbf{V}}

\newcommand{\vW}{\mathbf{W}}
\newcommand{\vx}{\mathbf{x}}

\newcommand{\vY}{\mathbf{Y}}

\newcommand{\vz}{\mathbf{z}}

\newcommand{\vbe}{\boldsymbol{\beta}}

\newcommand{\vnu}{\boldsymbol{\nu}}

\newcommand{\mbf}[1]{\mathbf{#1}}
\newcommand{\mbv}[1]{\mbox{\boldmath$#1$\unboldmath}}

\title{Bayesian Ensemble Echo State Networks for Enhancing Binary Stochastic Cellular Automata
}

\author{
  Nicholas Grieshop \\ Department of Statistics\\
  University of Missouri \\
  Columbia, MO \\
  \texttt{njgrieshop@mail.missouri.edu} \\
   \And
  Christopher K. Wikle \\ Department of Statistics \\
  University of Missouri \\
  Columbia, MO \\
}

\begin{document}

\maketitle

\begin{abstract}
    Binary spatio-temporal data are common in many application areas. Such data can be considered from many perspectives, including via deterministic or stochastic cellular automata, where local rules govern the transition probabilities that describe the evolution of the 0 and 1 states across space and time. One implementation of a stochastic cellular automata for such data is with a spatio-temporal generalized linear model (or mixed model), with the local rule covariates being included in the transformed mean response.  However, in real world applications, we seldom have a complete understanding of the local rules and it is helpful to augment the transformed linear predictor with a latent spatio-temporal dynamic process. Here, we demonstrate for the first time that an echo state network (ESN) latent process can be used to enhance the local rule covariates.  We implement this in a hierarchical Bayesian framework with regularized horseshoe priors on the ESN output weight matrices, which extends the ESN literature as well. Finally, we gain added expressiveness from the ESNs by considering an ensemble of ESN reservoirs, which we accommodate through model averaging. This is also new to the ESN literature. We demonstrate our methodology on a simulated process in which we assume we do not know all of the local CA rules, as well as a fire evolution data set, and data describing the spread of raccoon rabies in Connecticut, USA.
\end{abstract}

\keywords{ \and model averaging \and deep learning \and uncertainty quantification \and spatio-temporal \and dynamics \and reservoir computing}

\section{Introduction} \label{sec::intro}

Binary spatio-temporal data are common in many real-world data contexts, such as modeling the change in occupancy (presence/absence) of wildlife on a landscape \citep[][]{royle2007bayesian,broms2016dynamic,bertassello2021dynamic}, the spread of disease or invasive species \citep[][]{zhu2008autologistic,hooten2010statistical}, and the evolution of the boundary of a wildfire front \citep[][]{bradley2023deep}, to name just a few.  These types of data are often (but not always) gridded either naturally (e.g., satellite observations) or for convenience (wildfire modeling), in which case, each grid cell in the spatial domain of interest can be labeled with a 1 (presence) or 0 (absence). Models for such processes need to specify or learn some mechanism for the spatial field of 1s and 0s to change through time dynamically.  

Traditionally, one can model such processes in several ways. For example, building off the generalized linear mixed model (GLMM) time series literature, one can consider the spatio-temporal data to follow an independent non-Gaussian (Bernoulli) distribution conditioned on a latent Gaussian dynamic process \cite[e.g.,][]{west1985dynamic, gamerman1998markov,lopes2011generalized, cressie2011statistics}.  Another option is to consider such data as a binary Markov random field (i.e., an auto-logistic model; \citet{besag1972nearest, zhu2005modeling, zhu2008autologistic}).  Yet another approach considers the data to follow a cellular-automata (CA) with binary states with simple evolution rules that describe the change of the states over time \citep[e.g.,][]{ hooten2010statistical, hooten2020statistical}.  Note that there are overlaps between these various approaches as discussed in \citet[][]{wikle2015hierarchical}.  For example, one way to implement a stochastic CA model for binary data is to assume a conditionally independent Bernoulli data distribution as in spatio-temporal GLMMs, but consider local rules, informed by data, to describe the transition probabilities of the transformed mean response.  This is the general approach that we extend here. 

As summarized in \citet[][]{banks2021statistical}, it can be computationally challenging to estimate transition rules for CA models, which has somewhat limited their use in statistical applications.  This is particularly true when the rules are only partially known – e.g., known up to some parameters, or the more extreme case where we lack knowledge of the relevant class of rules.  Our interest here is to develop a general methodology for binary stochastic CA processes for dynamic spatio-temporal data that can learn the importance of various transition rules, but also account for unspecified, potentially nonlinear, latent dynamics that may control the CA transition probabilities. Unlike most deterministic CA implementations, we also require that this model  be embedded within a framework that can realistically account for the uncertainty associated with inference and predictions.

In recent years, efficient reservoir-based neural models such as echo state networks (ESNs) have shown a remarkable ability to account for unspecified dynamics in spatio-temporal data \citep[e.g.,][]{ Bianchi2015Short-TermDecomposition, mcdermott2017ensemble, mcdermott2019deep, bonas2021calibration, huang2022forecasting, yoo2023using}.
An ESN is a type of reservoir computing approach that is special a case of recurrent neural network \citep{Lukosevicius2009ReservoirTraining}.
The interconnecting neural network weights are not learned in an ESN but are randomly generated, and only the output weights are learned.  Remarkably, ESNs are still universal approximators \citep[][]{grigoryeva2018echo} and the sparsely interconnected hidden layers accommodate non-linear spatio-temporal behavior. In this sense, the ESN can provide a robust model for the latent dynamics in a binary CA model.

One of the drawbacks with ESNs, as with traditional recurrent neural networks, is that they do not provide a model-based estimate of uncertainty. The typical approach to quantify uncertainty is to use ensembles of (multiple) ESNs (based on different randomly generated internal weights) and to calibrate those to a desired coverage \citep[e.g.,][]{bonas2021calibration,yoo2023using}.  A few papers have tried to apply the concepts of Bayesian inference to the ESN, with varying degrees of rigor.  For example,
\citet{li2012chaotic,li2015echo} model the output from a Laplace distribution within a Bayesian framework. The output weights were given normally distributed prior distributions, and the resulting posterior distribution was maximized with a surrogate objective function.  Although this is Bayesian in the sense that it includes a data likelihood and priors, it does not utilize the full strength of Bayesian models as draws from the posterior are not uesd for uncertainty quantification of the resulting predictions.  On the other hand, \citet{mcdermott2019deep} considered ensembles of deep ESNs and then used a Bayesian stochastic search variable selection prior on the output weights associated with all of the ensembles of hidden units. Their work was in the context of continuous responses and not binary data or CA models as are our interests here.

In this paper, the focus is on spatially gridded binary response data that varies over time and that is modeled through a stochastic CA model. The primary novelty is the consideration of an ensemble of spatio-temporal ESNs to augment learning of the CA transition probabilities within a larger Bayesian hierarchical model in which output weight learning can be incorporated alongside traditional statistical modeling techniques. 
Uncertainty quantification of the resulting forecasts is provided via the Bayesian estimation and through a model averaging framework, which has not been applied to ESN-based models in the past. In addition, we use an efficient approach to obtain ESN tuning parameter ranges for use in the fully Bayesian model.
Section \ref{sec:Background} provides background details on binary dynamic spatio-temporal models, continuous ESNs, and binary ESNs.  Section \ref{sec:BinaryData} presents the methodology for our hybrid CA model with ESN dynamics, including implementation details.  This is followed by an evaluation of model performance on simulated data in Section \ref{sec:Sim} and on two real-world data sets, one modleing the spread of a fire front in a controlled burn experiment \ref{sec:burn}, and the other corresponding to the spread of raccoon rabies in Connecticut in Section \ref{sec:rabies}. Section \ref{sec:Conclusion} provides a brief conclusion. 

\section{Background } \label{sec:Background}
This section provides general background on spatio-temporal models for gridded binary dynamic spatio-temporal models used for CA applications and some useful background details on ESNs.

\subsection{Binary Dynamic Spatio-Temporal Models } \label{sec:BDTSMback}

Consider observations $y_{it}$ at spatial grid cells $\{ i=1,\ldots,n\}$ and discrete times $\{t=1,\ldots,T\}$ that follow a Bernoulli distribution,
\begin{equation}
y_{it} | p_{it} \sim \; indep. \; Bern(p_{it}),
\label{eq:BernData}
\end{equation}

\noindent where the transition probabilities $p_{it}$ control the stochastic CA evolution.  In a traditional binary stochastic CA model these probabilities are based on covariates ${\mbf x}_{it} \equiv (x^{(1)}_{it},\ldots,x^{(n_x)}_{it})'$ that are the $n_x$ potential local ``rules'' for the CA transition.  For example, a covariate might correspond to the number of neighbors in a queen's neighborhood of the $i$th cell that are in state 1, or the covariate may correspond to some spatial environmental variable such as elevation or proximity to a landscape feature such as a river. In this case, a simple spatio-temporal generalized linear model (GLM) can be used and one can consider the transformed mean response as
\begin{equation}
g(p_{it}) = {\mbf x}_{it}' {\mbv \beta},
\label{eq:glm}
\end{equation}

\noindent where the link function $g( \cdot)$ can be any of the usual Bernoulli link functions (e.g., logit, probit) and the $n_x$-dimensional parameter vector ${\mbv \beta}$ is estimated either with frequentist methods or via Bayesian methods with an appropriate prior distribution on ${\mbv \beta}$. These estimates then suggest which of the ``rules'' are most important for the CA evolution.

Typically, in real-world applications, one does not know all of the rules that could be important to describe a particular stochastic CA evolution. Then, as in spatio-temporal GLMMs \citep[e.g.,][]{cressie2011statistics}, one can consider a latent dynamic process as a surrogate for the unknown rules,
\begin{equation}
    g(p_{it}) = {\mbf x}_{it}' {\mbv \beta} + {\mbf v}'_{i} {\mbf h}_t,
\end{equation}
where ${\mbf h}_t$ is a $n_h$-dimensional latent dynamic process (typically either $n_h=n$-dimensional corresponding to all spatial locations, or $n_h < n$-dimensional, corresponding to a reduced rank dynamic process).  In these cases,  ${\mbf v}_{i}$ is either the identity or a set of known spatial basis functions. One can then model ${\mbf h}_t$ as a discrete-time dynamic process such as a vector autoregression or quadratic nonlinear model with Gaussian errors \citep[see][for more details]{cressie2011statistics}.  As described below, here we turn this around and will generate ${\mbf h}_t$ via an ESN and then estimate ${\mbf v}_i$ with Bayesian regularization.

\subsection{Echo State Networks} \label{sec:ESNback}
A basic vanilla ESN \citep[][]{lukosevicious} applied to a Gaussian output response at $n$ spatial locations, $\vO_t = (o_{1t} \dots o_{nt})'$ and times $t = 1, \dots, T+1$ with observed $n_z$-dimensional input vectors $\vz_t$,  can be written as 

\begin{align} 
\text{response}: &\quad {\mbf O_t}={\mbf V}{\mbf h}_t \nonumber \\
\text{hidden state}:
&\quad {\mbf h}_t =g_h \bigg(\frac{\nu}{|\lambda_w|}{\mbf W}{\mbf h}_{t-1} + {\mbf U}{\mbf z}_t\bigg), \label{eq:h}\\
\text{parameters}:  &\quad  {\mbf W}=[w_{i,\ell}]_{i,\ell}:\gamma_{i,\ell}^{w}Unif(-a_w,a_w)+(1-\gamma_{i,\ell}^{w})\delta_0 \label{eq:Weq} \\
&\quad {\mbf U}=[u_{i,j}]_{i,j}:\gamma_{i,j}^{u}Unif(-a_u,a_u)+(1-\gamma_{i,j}^{u})\delta_0  \label{eq:Ueq}\\
&\quad \gamma_{i,\ell}^{w} \sim Bern(\pi_w), \;\; \gamma_{i,j}^{u} \sim Bern(\pi_u), \label{eq:gammas}
\end{align}

\noindent where the $n_h$-vector ${\mbf h}_t$ corresponds to the hidden units, and the $n_h \times n_h$ matrix ${\mbf W}$ and $n_h \times n_z$ matrix ${\mbf U}$ are weight matrices with elements drawn randomly from the specified distributions in (\ref{eq:Weq}) and (\ref{eq:Ueq}) with hyperparameters $a_w$, $a_u$, $\pi_w$ and $\pi_u$. Here, $\delta_0$ is a Dirac delta function at zero and the  hyperparameters $\pi_w$ and $\pi_u$ determine the sparsity of the matrices ${\mbf W}$ and ${\mbf U}$, respectively. Furthermore, $\lambda_w$ is the spectral radius of ${\mbf W}$ and $\nu$ is a tuning parameter that controls the ``echo state property,'' which corresponds to how sensitive the hidden units are to the initial conditions. The element-wise activation function $g_h(\cdot)$ is typically a $tanh(\cdot)$. Note that the role of the ESN hidden units is to nonlinearly and randomly transform the inputs into a higher dimensional space and to simultaneously remember the input \citep[e.g.,][]{lukosevicious}. The sequence $\{ \vh_1 \dots \vh_T \}$ is sometimes said to be a reservoir.

Importantly, the elements of the matrix ${\mbf V}$ are output weights that are learned via regularization, typically through a ridge penalty assuming independent additive errors, ${\mbf O}_t = {\mbf V}{\mbf h}_t + {\mbv \epsilon}_t$. Note that in many cases, the inputs, $\vz_t$, correspond to the response at previous time steps, but one can also include exogenous covariates in addition to, or in place of the lagged inputs. The choice of inputs is application-specific.


The ESN is critically dependent on the fixed but random generation of the sparse matrices $\vW$ and $\vU$, which is both a benefit and a curse. On the one hand, the ESN does not need to learn these weight matrices, which allows for it to be estimated rapidly and used with much less training data than a traditional recurrent neural network (RNN). On the other hand,
the overall size of the ESN hidden layer ($n_h$) will, on the whole, be larger than a traditional RNN to achieve the same performance \citep{prokhorov2005echo}, and the output can be sensitive to the particular reservoir weights that are randomly selected.  This vanilla ESN can be extended in many ways, such as the leaky integrator ESN \citep{jaeger2007optimization}, where updates to the reservoir can include heavier weighting of nearby observations. Deep versions of ESNs 
have also been considered \citep[e.g.,][]{ma2017deep,mcdermott2019deep}.

To create uncertainty for the resulting predictions, multiple reservoirs can be constructed by iterating through the same procedure and resampling the components of $\vW$ and $\vU$ to create an ensemble of models and forecasts. This ensemble approach, as it applies to neural networks, is a well-known technique for accounting for uncertainty \citep[e.g.,][]{hashem1995}. In the context of ESNs, the ensemble approach has been used in several studies \citep[e.g.,][]{yao2013,mcdermott2017ensemble, rigamonti2018ensemble, yoo2023using}.
In some instances, the ensembles are weighted equally, whereas in others, the model weights are adjusted to calibrate the forecast intervals. In either case, the repeated creation and subsequent prediction of multiple ESNs often gives a plausible range of prediction values and uncertainty quantification. One can also develop uncertainty quantification for ESNs  by borrowing the idea of dropout \citep{srivastava2014dropout}. For example, 
\citet{Atencia} consider multiple reservoirs that are constructed by fitting a traditional ESN and then zeroing out some of the members and utilizing the same procedure as the ensemble methods to develop confidence intervals for the predictions. Lastly, as mentioned in the Introduction, Bayesian approaches have been implemented for ESNs \citep[e.g.,][]{li2012chaotic,li2015echo,mcdermott2019deep}, although to date, no one has implemented Bayesian shrinkage approaches for ESNs with binary observations.

\subsection{Binary ESNs}\label{sec:BinaryESN}
ESNs have been applied to binary observations in the context of classification. An example of this is the experiments considered in \cite{Jaeger2012LongStudy}. However, the output weights in those examples are trained in a continuous ridge regression framework with the binary data treated as if continuous, and then the model predictions (which are continuous) are converted to 0 and 1 through a thresholding procedure. Other applications of binary data include \cite{Yilmaz2014ReservoirAutomata} and \cite{Nichele2017DeepAutomata}, where both the input and output are binary, but in this special case the reservoir weights are cellular automata rules.
Again, the final output weights are trained via linear regression with thresholding. Both methods, with real valued inputs or binary inputs, succeeded in the 5-bit memory test \citep{jaeger2007optimization}, which demonstrates the ESN's long-term memory property.
In these binary applications, it was demonstrated that by varying the cellular automata update rule, i.e. $\vU$, the different outputs had differing levels of success, which could then be used to quantify uncertainty. 

In these examples, the classification problem was treated as a regression problem. Although not ideal from a statistical modeling perspective because treating binary data as continuous violates the implicit normality assumption in the errors, this approach does lead to large computational savings, as ridge regression or Moore-Penrose generalized inverse implementations are relatively computationally simple and efficient to apply.  From a statistical perspective, it is more appropriate to consider binary ESNs from the perspective of regularized GLMs, either in a frequentist or Bayesian implementation.  
For example, the output weights were trained via a logistic regression in \cite{pascanu2015malware} and with a support vector machine classifier in \cite{scardapane2017semi}.  In the next section we describe a Bayesian approach that embeds the ESN within a binary spatio-temporal CA for a given set of reservoir weights, and treats the ensemble of such outputs through a model-averaging perspective.

\section{ESN-Enhanced Bayesian Binary CA Model } \label{sec:BinaryData}

Here we describe a novel model for binary spatio-temporal data that augments local covariate rules in the CA transition probabilities with ESN reservoirs. This is implemented within a hierarchical Bayesian framework and utilzes model averaging to account variation in the ESN reservoirs. 

\subsection{Bayesian Binary CA Response Model} \label{sec:BinBayes}

As described in \ref{eq:BernData} we assume the binary data, $\{y_{it}: i=1,\ldots,n; t=1,\ldots,T\}$, follow an independent Bernoulli distribution, conditioned on the transition probabilities. Let ${\mbf Y}_t = (y_{1t},\ldots,y_{nt})'$ and ${\mbf p}_t = (p_{1t},\ldots,p_{nt})'$. Then, 
\begin{equation}
{\mbf Y}_t | {\mbf p}_t \sim Bern({\mbf p}_t),
\label{eq:Bern_vec}
\end{equation}
where 
\begin{equation}
logit({\mbf p}_t) = {\mbv \alpha} + {\mbf X}_t {\mbv \beta} + {\mbf V} {\mbf h}_t,
\end{equation}
and ${\mbv \alpha}$ is an $n$-dimensional offset vector, ${\mbf X}_t$ is an $n \times n_x$ matrix of local covariates, ${\mbv \beta}$ is an $n_x$-vector of coefficients, ${\mbf V}$ is an $n \times n_h$ matrix of ESN output coefficients, and ${\mbf h}_t$ corresponds to a reservoir from an ESN. The model is completed by utilizing a regularized horseshoe prior  \citep{piironen2017sparsity} for the elements of the output matrix, $\vV$ (denoted, $V[i,j]$), and relatively vague conjugate priors for ${\mbv \alpha}$ and ${\mbv \beta}$. Specifically, 
\begin{align}
    V[i,j] & \sim N(0,1)\tau \Tilde{\lambda}_{[i,j]} \\
     \Tilde{\lambda}_{[i,j]} & = \frac{c^2 \lambda_{i,j}^2}{c^2 + \tau^2 \lambda_{i,j}^2} \label{eq:priorLam} \\
     c & = scale_{slab} \sqrt{c_{aux}} \\
     [\lambda_{i,j}] & \sim \textit{half-t}(\nu_{local}, 0, 1) \\
     [c_{aux}] & \sim InvGam(df_{slab} / 2,df_{slab} / 2) \\
     [\tau] & \sim \textit{half-t}(\nu_{global}, 0, 2 scale_{global}) \\
     {\mbv \alpha} & \sim N({\mbf 0},\sigma^2_\alpha {\mbf I}) \\
     {\mbv \beta} & \sim N({\mbf 0},\sigma^2_\beta {\mbf I}).
\end{align}
The advantage of utilizing the regularized horseshoe prior over the horseshoe prior of \cite{carvalho09a} is that the regularized version shrinks all coefficients, and this property avoids some issues that could arise from binary data with fully separable data.
Under this construction, in Equation \ref{eq:priorLam} if a value of $V[i,j]$ is close to zero, then $c^2 > \tau^2 \lambda_{i,j}^2$ and $\Tilde{\lambda}_{i,j} \approx \lambda$, which is the original horseshoe prior.
 When elements of $V[i,j]$ are non-zero then  $c^2 < \tau^2 \lambda_{i,j}^2$ and $\Tilde{\lambda}_{i,j} \approx c^2 / \tau^2$ and therefore will regularize non-zero coefficients. Note, additional shrinkage priors could be assigned to the covariate parameters ${\mbv \beta}$ if there are a large number of potential covariates being considered.
 
 Given posterior samples of $\vV$, ${\mbv \alpha}$, and ${\mbv \beta}$, posterior predictive distributions of forecasts can be obtained by calculating ${\mbf p}_{T+1}$, given known values of ${\mbf X}_{T+1}$ and ${\mbf h}_{T+1}$. We assume that we have the local covariates at time $T+1$ and can get ${\mbf h}_{T+1}$ by continuing the ESN iterations.
The ESN reservoirs are obtained from Equations \ref{eq:h} - \ref{eq:gammas}, with inputs ${\mbf z}_t$, which are application specific. As discussed in Section \ref{sub:label} below, we select $K$ different sets of reservoirs based on different draws of the random elements in the reservoir weight matrices and perform the Bayesian inference $K$ times. This procedure lends itself easily to parallel computing as all models can be fit independently at this stage.

\subsection{Implementation Details} \label{sec:ImplementationDetails}

Conditional on the ESN reservoir sequence, the model presented in Section \ref{sec:BinBayes} is simply a spatio-temporal Bayesian logistic regression with regularized horseshoe priors on the ESN reservoir output weights, and with random spatial intercepts and covariate parameters.  The model was implemented using RStan \citep{rstan} (see Appendix \ref{ap:stan} for an example of the Stan code for our model).

There are two important implementation issues that arise due to the use of the ESN reservoirs in the model. The first has to do with the selection of prior distributions for the ESN tuning parameters. The second has to do with how we utilize ensembles of different ESN reservoirs within our Bayesian model. Sections \ref{sub:tune} and \ref{sub:label} describe our approaches to these problems, respectively, which is novel in the ESN and CA literature.

\subsubsection{Tuning Parameter Selection} \label{sub:tune}
As shown in Section \ref{sec:ESNback}, there numerous tuning parameters that must be selected for ESNs. This is often done through a validation/cross-validation framework, which is tractable in the context of continuous responses where ridge-regression is typically used to estimate the output weights. As discussed in Section \ref{sec:BinaryESN}, for the sake of reduced computation time it is common to treat the binary response in ESNs as if it were a continuous variable and use ridge-regression estimation for regularized inference \citep{jaeger2007optimization}.  Although our model includes a more realistic binary likelihood, we use this approach to obtain plausible (empirical Bayesian) prior distributions for the tuning parameters
 ($\vnu$, $\boldsymbol{\pi}_{w}$, $\boldsymbol{\pi}_{u}$, $\boldsymbol{a}_w$, $\boldsymbol{a}_u$, $\boldsymbol{n}_{h}$). This is the first time such a procedure has been used in the context of neural Bayesian computing and ESNs.  In particular, a grid search is conducted over a large range of plausible parameter values to determine which values result in good predictions. In particular, each tuning parameter is assigned a range of plausible values, and an exhaustive fit of the approximate ridge regression on the output weights applied to (assumed continuous) responses is performed over all $C$ combinations. 
The mean squared error is computed for each of these $C$ combinations, and the top 100 parameter combinations are recorded. From these 100 best combinations, the $10^{th}$ and $90^{th}$ percentile is computed for each parameter, which are then used as the lower and upper limit of the parameter range prior distributions.
This produces a range of plausible values, in terms of minimizing mean squared error, to use in our Bayesian inference procedure. These ranges suggest ranges on uniform distributions that are used as prior distributions in the more rigorous Bayesian binary model inference procedure described in Section \ref{sec:BinBayes}. This algorithm is given in Algorithm \ref{algo:tuning}.

\begin{algorithm}[H]
\caption{Searching for tuning parameter ranges}
\begin{algorithmic}[1]
\algblock{Input}{EndInput}
\algnotext{EndInput}
\algblock{Output}{EndOutput}
\algnotext{EndOutput}
\newcommand{\Desc}[2]{\State \makebox[2em][l]{#1}#2}
\Input \, $\{\vY_t$, $\vx_t$\} for $t=1,\ldots,T$; plausible set of C ESN reservoir parameters: $\vnu$, $\boldsymbol{\pi}_{w}$, $\boldsymbol{\pi}_{u}$, $\boldsymbol{a}_w$, $\boldsymbol{a}_u$, $\boldsymbol{n}_{h}$  \EndInput
\For {c $\in$ 1, 2, \ldots, C}
    \State Select the $c$th element of  $\vnu$, $\boldsymbol{\pi}_{w}$, $\boldsymbol{\pi}_{u}$, $\boldsymbol{a}_w$, $\boldsymbol{a}_u$, $\boldsymbol{n}_{h}$
    \State Compute $\vh^c_t$ for $t=0,\ldots,T-1$ using Equations (\ref{eq:h}) - (\ref{eq:gammas}), with inputs $\vx_t$ and parameters from step 3.
    \State Compute $\vV^c = \vY \vH^{c'} \left(\vH^c \vH^{c'} + diag(\lambda) \right)^{-1}$, where $\vY \equiv [\vY_1,\ldots,\vY_{T-1}]$, $\vH^c \equiv [\vh^c_1,\ldots,\vh^c_{T-1}]$. 
    \State Compute the out-of-sample ESN reservoir for prediction, $\vh^c_{T}$, from the ESN equations
    \State Obtain forecast, $\widehat{\vY}^c_{T} = \vV^c \vh^c_{T}$
    \State Compute MSE using $\vY_{T}$ and $\widehat{\vY}^c_{T}$
\EndFor
\State From the top $100 < C$ candidate models based on minimum MSE, compute the $10^{th}$ and $90^{th}$ percentile for each parameter and used these values as limits to a uniform prior distribution for that parameter. 
\end{algorithmic}
\label{algo:tuning}
\end{algorithm}

\subsubsection{Model Weighting}
\label{sub:label}

Although ESNs are, in principle, universal approximators, in practice their predictive performance can be sensitive to the particular reservoir, suggesting that multiple ESNs (ensembles) should be used, where each ESN arises from a different draw of the reservoir weight matrices $\vW$ and $\vU$  \citep[e.g., see ][]{mcdermott2017ensemble}. 
From Section \ref{sub:tune}, plausible distributions for the tuning parameters that control these reservoir weight matrices can be determined and one can generate $K$ different sets of reservoirs, $\vH^{(1)},\ldots,\vH^{(K)}$, where $\vH^{(k)} \equiv [\vh_1^{(k)},\ldots,\vh_T^{(k)}]$ corresponds to the $n_h \times T$ reservoir matrix for the $k$th ensemble. One can then fit the Bayesian model in Section \ref{sec:BinBayes} for each of the $K$ reservoirs.  Note, these $\vH$ matrices are different than those used in Algorithm \ref{algo:tuning}.

Rather than select one of the models from the different reservoirs, we perform model averaging (multi-model inference or ensemble learning)  \citep[e.g.,][]{hoeting1999bayesian, anderson2002avoiding, dong2020survey}. Such model averaging procedures are particularly beneficial when the focus is on prediction, but when the focus is on inference (say, describing which covariates are most useful), it can be difficult to summarize the influence of a particular variable across all model fits. It is also necessarily more computationally expensive to fit multiple models and it is more difficult to do model diagnostic checking \citep[e.g.,][]{ver2015iterating}. However, our focus here is on prediction (forecasting) and our multiple ensemble ESN reservoir calculations and Bayesian fitting are embarrassingly parallel, so we have found the additional model expressiveness worth the computational and diagnostic cost.

To implement our model averaging, we compute the output weights and coefficients for each of the $K$ unique models, ${\mbf V}^{(k)}$, ${\mbv \alpha}^{(k)}$, and $\vbe^{(k)}$ using the procedure in Section \ref{sec:BinBayes}.
Then, we obtain the state probability at location $i$ at time point $t$ from the $k^{\mathrm{th}}$ model, denoted by $p_{it}^{(k)}$ and the truth is denoted by $y_{it}$.
Thus to compute model weights, we can use a simple weighting process to find the weights $w_1,\ldots,w_K$ that minimize:
\begin{equation}
    \mathrm{Loss} = -\frac{1}{nT} \sum_{i=1}^{n} \sum_{t = 1}^T y_{it} \cdot \mathrm{log}(w_1 p_{it}^{(1)} + \dots + w_K p_{it}^{(K)}) + (1 - y_{i,t}) \cdot \mathrm{log}(1 - (w_1 p_{it}^{(1)} + \dots + w_K p_{it}^{(K)})),
\end{equation}
subject to the constraint that $w_k \geq 0$ and $\sum w_k = 1$.
These constraints ensure that each of the $k$ members can only add positively to the final weighted model. 
The weights were found using the method of \citet{byrd1995limited} as implemented in base R \citep{Rstuff}. Note, that an alternative to this weighting approach would be to use formal Bayesian model averaging (BMA) \citep{hoeting1999bayesian}. 
However, BMA is known to be computationally expensive, and the simplified model weighting scheme presented here has less computational cost and fits naturally with our embarrassingly parallel implementation. 

This procedure of weighting multiple models allows for information from the more informative reservoirs to be favored - as these reservoirs are randomly generated, combining them negates some of the variability that is inherently present from the random reservoir weight matrices.
This weighting procedure is an extension of the typical ensemble treatment of uncertainty quantification in ESN where each member contributes equally to the final estimates \citep[e.g.,][]{mcdermott2017ensemble}.
In addition, learning the output weights via a Bayesian hierarchical model allows there to be uncertainty in the model predictions.
For every posterior draw of the model parameters, the weighted probability of transition can be computed.
Thus, from these weighted probabilities the highest posterior density (HPD) can be computed, which captures the uncertainty of the predicted transition probability.

\section{Simulation Experiment and Applications}

We demonstrate the model with a binary CA simulation model in Section \ref{sec:Sim}. This is followed by an application to an experimental fire burn data set in Section \ref{sec:burn} and the spread of raccoon rabies in Connecticut in Section \ref{sec:rabies}.

\subsection{Stochastic CA Simulation}
\label{sec:Sim}

\begin{figure}[H]
    \centering
    \includegraphics[width = 0.9\textwidth]{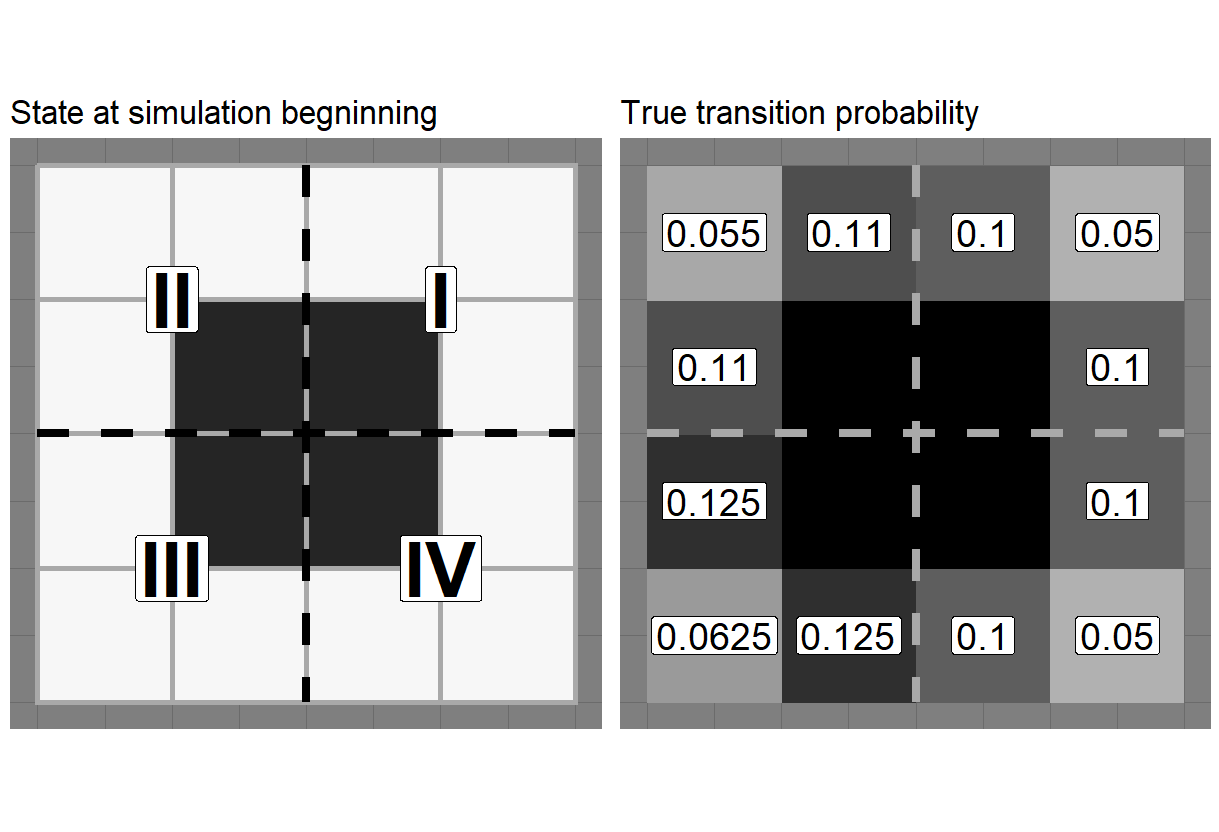}
    \caption{The left panel shows a small portion of the domain at the start of the simulation. The Roman numerals indicate the quadrant divisions relative to each $4 \times 4$ set of grid locations. The right panel demonstrates how the transition probability is a function of the number of neighbors and the location of the cell - quadrant I and IV share the same true transition probabilities whereas quadrants II and III are unique.}
    \label{fig:simDat}
\end{figure}

\begin{figure}[H]
    \centering
    \includegraphics[width = 0.7\textwidth]{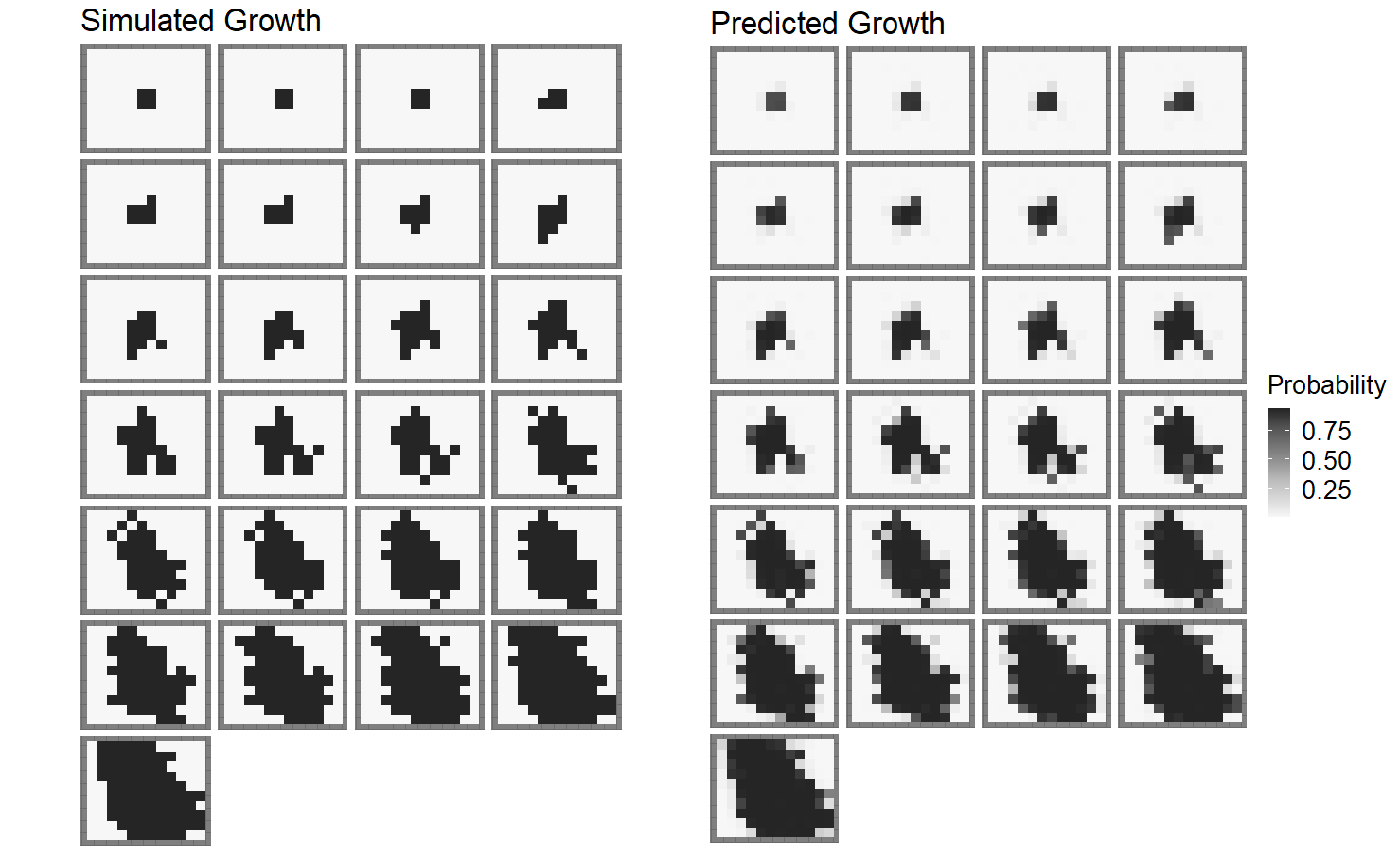}
    \caption{The true states of the simulation for the first 25 times are presented on the left side of the figure where a black cell corresponds to state 1 (presence) and a white cell corresponds to state 0 (absence). The time sequence is from left to right and top to bottom. The mean in-sample probabilities of transition for our ESN-enhanced binary CA model are presented on the right, demonstrating the ability of our model to capture the evolution of the true process.}
    \label{fig:simTruth}
\end{figure}

A simulation was constructed to illustrate our methodology applied to binary stochastic CA prediction.
The simulation considers a diffusive process evolving over 26 times on  a $10 \times 12$ regular spatial grid,  with the first 25 time points used for training purposes.
At the initial time step the four center grid cells were the only cells assigned to state 1 (i.e., presence). 
Cell transitions were assumed to be a function of the number of queen's neighbors that were of state 1, with a probability of transitioning between state 0 and 1 with one neighbor in state 1 being 5\%, two neighbors in state 1 being 10\%, and so forth.
Additionally, if a cell was in quadrant II or III (see the left panel in Figure \ref{fig:simDat}) it had an increased chance of transitioning, with cells in quadrant II transitioning from 0 to 1 with a 10\% increase in probability and cells in quadrant III with a 25\% increase in probability. These transition rules are summarized in the following two equations:
\begin{align}
    y_{it} | p_{it} & \sim Bernouilli(p_{it}) \\
    p_{it}  & = 0.05 x_{it} \left(1 + 0.1 I_{\text{Quad II}} + 0.25 I_{\text{Quad III}} \right), 
\end{align}
where $x_{it}$ is the total number of queen neighbors in state 1 at location $i$ and time $t$, $p_{it}$ is the associated probability of transitioning from state 0 to state 1, and $I_{\text{Quad II}}$ and $I_{\text{Quad III}}$ are indicators if the cell is in quadrant II or III, respectively. Note, it is assumed that once a cell is in state 1, it is ``frozen'' and cannot transition back to state 0.
The right panel in Figure \ref{fig:simDat} shows how this structure implies non-homogeneous transition probabilities. That is, the right side of the domain, quadrants I and IV, have the same transition probabilities and the left side of the domain, quadrants II and III, have probabilities that are different than those on the right side of the domain and different from each other (
quadrant III has a higher transition probability than any other quadrant).
Such unequal transition probabilities could arise in a real-world application where for example, in a fire spread model, certain cells could have particular fuel characteristics which increase cell transitions. 
The evolution of the simulated process is shown in the left panels of Figure \ref{fig:simTruth}. 

The main purpose of this simulation is to show that our ESN-enhanced binary CA model can adapt to a situation where our local rules (covariates) are not completely known - that is, our model is misspecified.  In this case, we assume we only have covariates that represent the number of neighbors in state 1 and a single indicator if the cell was on the left side (quadrant II or III) versus the right side (quadrant I or IV); thus, we do not know that quadrant II and III have different probabilities of transition and the ESN reservoir component must try to adapt.

\begin{figure}[H]
    \centering
    \includegraphics[width = 0.9\textwidth]{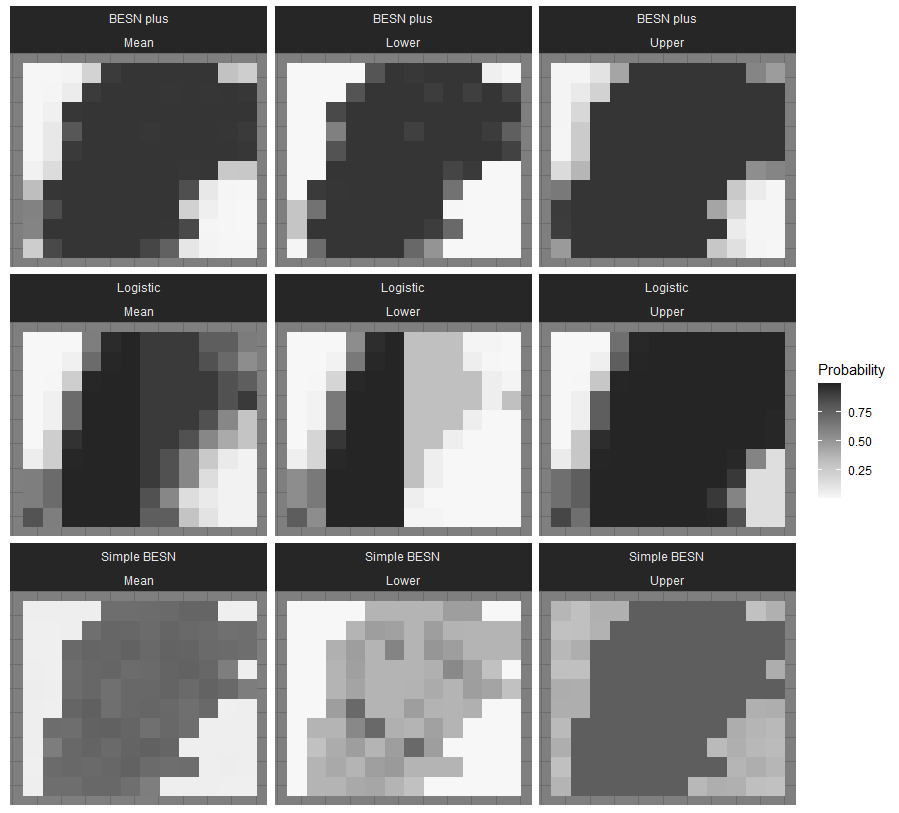}
    \caption{The comparison of the results for the three models applied to the simulation data. The first row shows results for the BESN-plus model whose lower and upper 95\% HPD intervals (second and third column) are much smaller than the other three methods. The second row shows the logistic model results, and the third row shows the results for a BESN without additional covariates, which has much wider HPD intervals than the the BESN-plus model.}
    \label{fig:simPred}
\end{figure}

Figure \ref{fig:simPred} shows the out-of-sample prediction for time 26 (the bottom figure in Figure \ref{fig:simDat}) using the first 25 times as training. 
The number of the queen's neighbors in state 1 was used as inputs to the ESN.
As mentioned above, the transition probabilities for this model considered the covariates associated with whether the cell was on the left or right side of the domain as well as the ESN reservoirs. The model captures the probabilities of state transition very well in the sense that it mimics the true spread very closely. 

For comparison, in addition to our model that includes the ESN and local covariates (which we call the Bayesian ESN plus covariates model, ``BESN-plus model'') we also fit the model with these same covariates but without the ESN reservoirs (which we call the ``logistic model''), and a third model without the covariates but with the ESN reservoirs (which we call the simple ``BESN model''). 
The BESN-plus model captured the growth and provided uncertainty for the probability of spread better than the other two as shown in Figure \ref{fig:simPred}.
Compared to the Bayesian logistic regression with the same covariates (the number of neighbors that are state 1 and indicator if the cell was in quadrant I or IV) the results were worse for the logistic model.
There was a similar improvement over the BESN constructed without any covariate information. 
This is to be expected because the inclusion of additional information should, at worst, have no improvement on the forecasts.

To more formally evaluate these model predictions, we consider the 
Brier score \citep{brier1950},
\begin{equation}
    BS = \frac{1}{N T} \sum_{t =1}^T \sum_{i=1}^{N} (p_{it} - o_{it})^2,
\end{equation}
were $i=1,\ldots,n$ correspond to spatial locations, $t=1,\ldots,T$ corresponds to times, $p_{it}$ is the predicted probability, and $o_{it}$ the  observation truth (0 or 1). A lower Brier score is better.
The resulting Brier scores were 0.0475 for the BESN-plus model, compared to a score of 0.1226 for the logistic model, and a score of 0.1179 for the BESN without covariates. This demonstrates the ensemble reservoir models' ability to improve model predictions for misspecified models and the improvement from a standard ESN without covariate information.

\subsection{Experimental Burn Fire Evolution}
\label{sec:burn}

\begin{figure}[H]
    \centering
    \includegraphics[width = 0.8\textwidth]{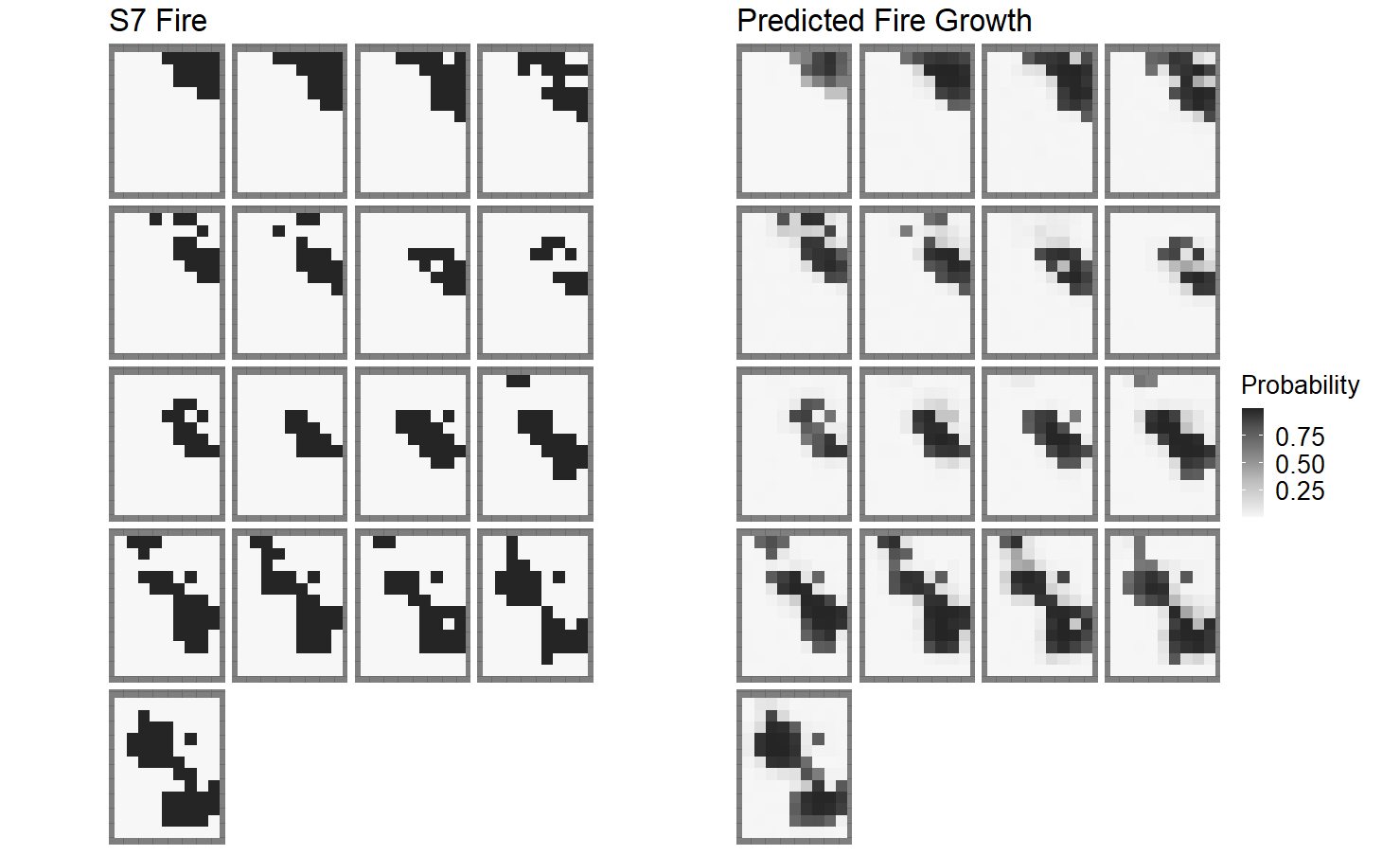}
    \caption{The true state of the first 17 time points of the S7 prescribe fire are presented on the left side of the figure. Each row is four time steps long, with the fire beginning in the top left. The right side of the figure shows the in-sample mean predictions from the BESN-plus method.}
    \label{fig:s7growth}
\end{figure}

We now consider data from the RXCadre series of experimental burns from Florida, USA, where an
infrared camera recorded the area's temperature during controlled fire burns, and local weather conditions were recorded \citep{RXcadreExperiment}.
The data came from the ``S7'' fire and were originally $240 \times 320$ resolution, but were averaged onto a $24 \times 32$ grid for use here, with the mean temperature of the higher-resolution cells within each lower-resolution cell cell used as data. 
The criterion for the state classification was based on the temperature of the cell, with progression from unburned to burning after the temperature crossed a threshold above the background average of 300K.
A cell's state was solely based on temperature so a cell was able to transition from a burning state back to non-burning state.
This can be seen in Figure \ref{fig:s7growth}, where the burning cells start in the top right of the domain and spread down and to the left.
As the fire spreads, the cells which were at one point burning transition to a non-burning state.
The first 17 time points were used to train the model.
The inputs to the ESN was the temperature value at the previous time point. 
In this application, we assumed the local covariates were the number of neighboring cells that were burning at the previous time point.
The results in Figure \ref{fig:s7growth} and Figure \ref{fig:simPred} demonstrate the ability of the BESN-plus model to capture the growth and movement of a real-world binary system.
This example also shows how the method is applicable to cellular spread models with non-terminal states - as it was able to capture the transition back to a non-burning state.

\begin{figure}[H]
    \centering
    \includegraphics[width = 0.5\textwidth]{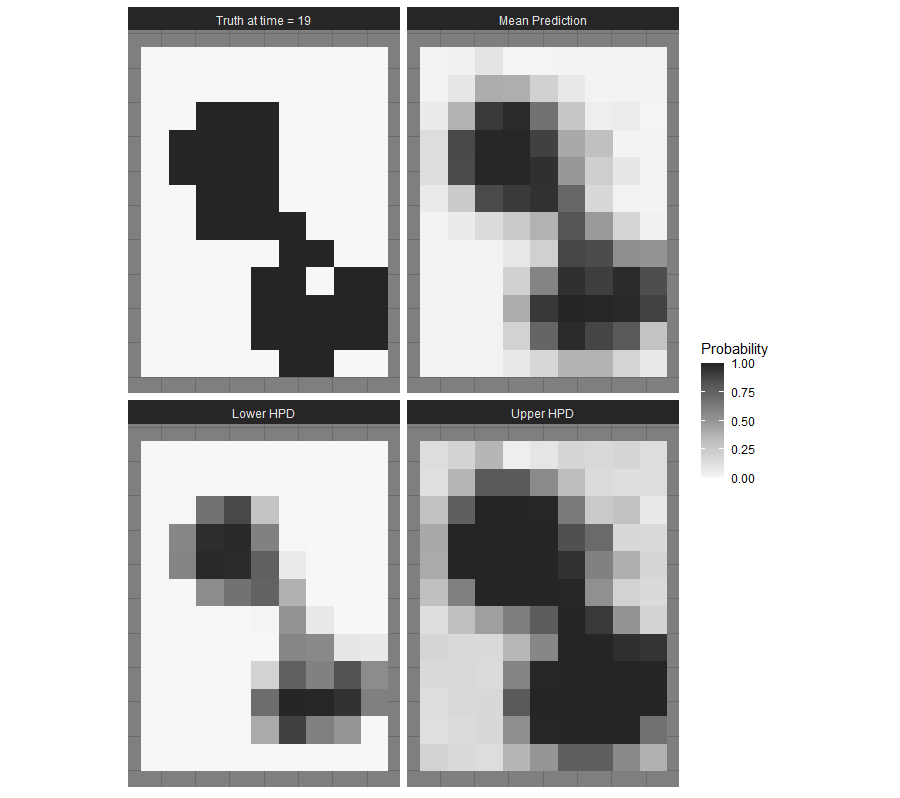}
    \caption{The true state of the S7 fire at time 18 is presented in the top left panel. The other three plots show the mean value from the BESN-plus method (upper right panel) along with the lower (lower left panel) and upper (lower right panel) bounds of the 95\% credible intervals.}
    \label{fig:s7}
\end{figure}

Figure \ref{fig:s7} shows the results for the out-of-sample prediction at time 18.
The BESN-plus methodology allows for full uncertainty quantification of future steps and this can be seen in Figure \ref{fig:s7} where the growth of the fire is predicted for the next time step. The credible intervals show good coverage in this case. 

\subsection{Spread of Raccoon Rabies in Connecticut}
\label{sec:rabies}
\begin{figure}[H]
    \centering
    \includegraphics[width = 0.7\textwidth]{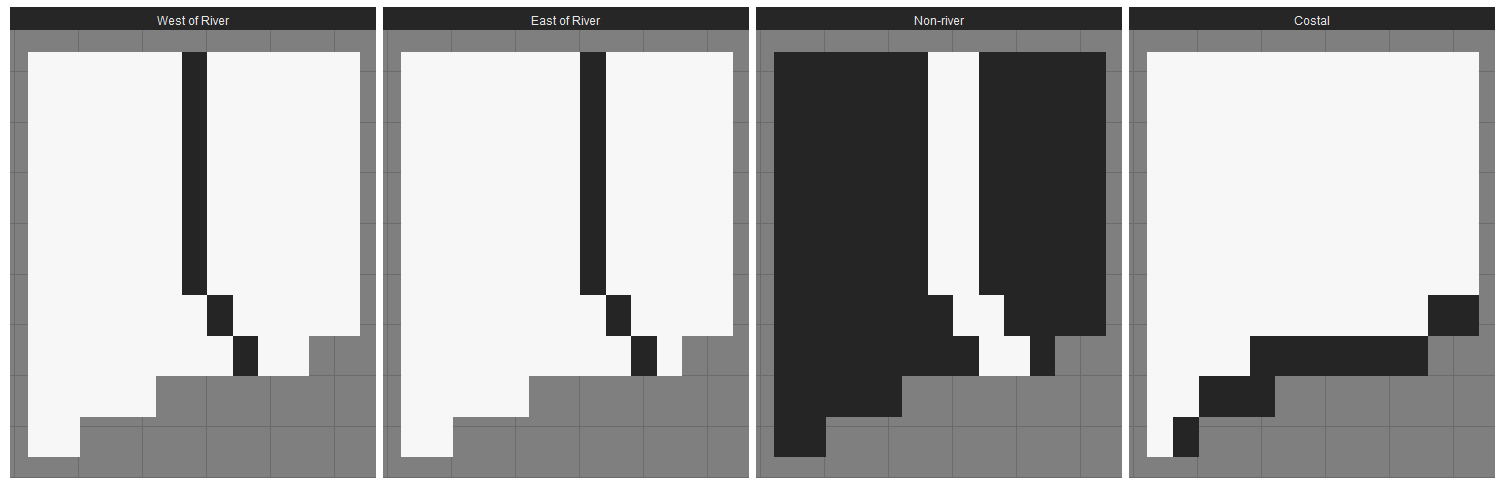}
    \caption{Plots showing the local covariate information from left - counties that neighbor the Connecticut River (west and east sides), non-river counties, and coastal counties. Note that the jagged southern edge is the Long Island Sound and the west, north, and eastern edges are the states of New York, Massachusetts, and Rhode Island, respectively.}
    \label{fig:rabiesCovNA}
\end{figure}

\begin{figure}[H]
    \centering
    \includegraphics[width = 0.8\textwidth]{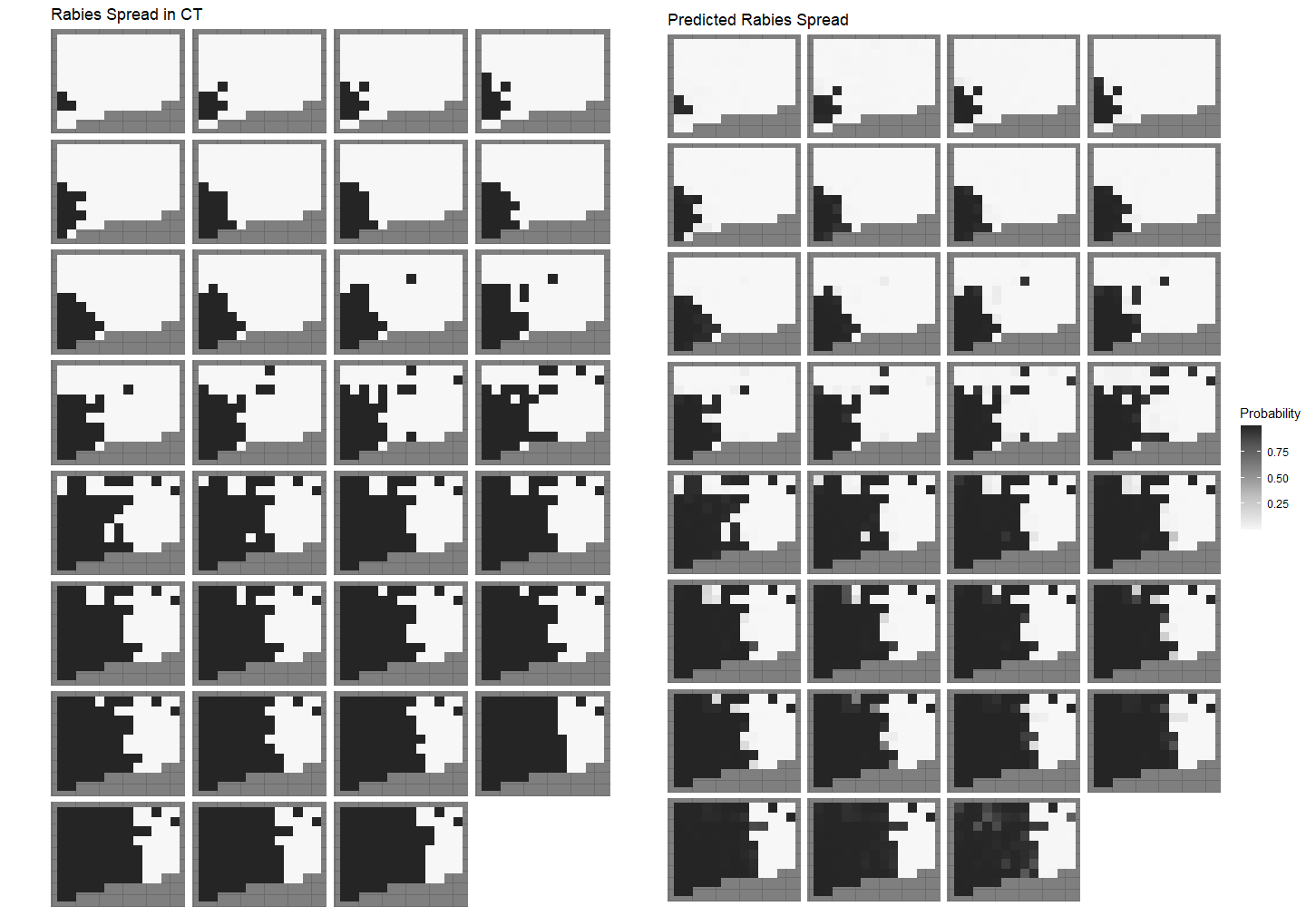}
    \caption{The left side of the figure shows the progression of the rabies virus in Connecticut. The figure reads from left to right and top to bottom, so the top row shows the first four-time steps, the second row the next four, and so on. The right side of the figure shows the mean of the transition probabilities at each time step. The inputs to the ESN model are the number of queen's neighbors with rabies present at the previous time step. The last panel (third column of the last row) on both the left and right figures is the {\color{red}the out-of-sample forecasted prediction probabilities. See Figure \ref{fig:CThpd} for details.}}
    \label{fig:CTspread}
\end{figure}

\begin{figure}[H]
    \centering
    \includegraphics[width = 0.7\textwidth]{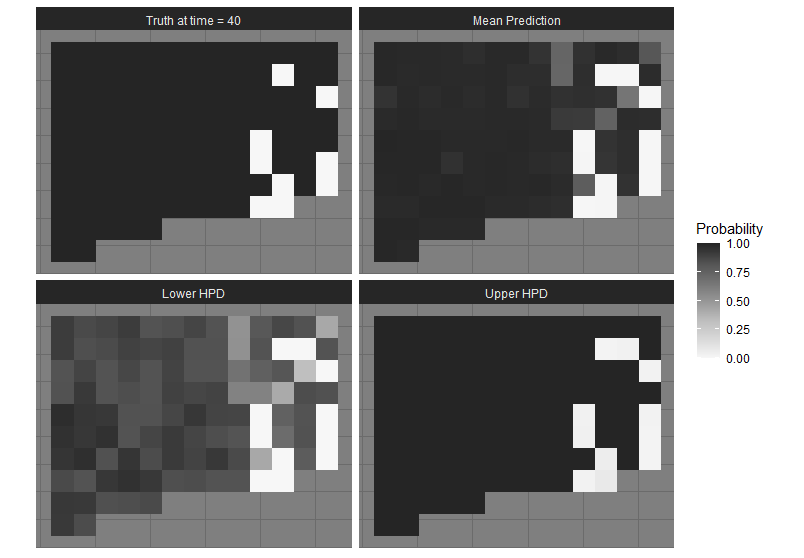}
    \caption{The true state of the rabies outbreak in Connecticut at time point {\color{red} 31} alongside the mean value from the BESN-plus posterior distribution. Presented on the bottom row are the lower and upper 95\% HPD intervals. 
    }
    \label{fig:CThpd}
\end{figure}

The Connecticut raccoon rabies dataset, as analyzed in \cite{smith2002predicting}, indicates when the first occurrence of rabies was discovered in the different counties of Connecticut. 
The rate of spatial propagation of the rabies virus was found to be slowed by the presence of rivers.
The original county-level data was fixed into regular grid cells as was done in \cite{hooten2010statistical}.
There are 109 regularly gridded cells approximating the counties in Connecticut, USA. 
The rabies data were collected over 48 months.
The data used to fit the model was constrained to the first 30 time steps.
The number of counties in a queen's neighborhood with recorded rabies was used as input into the ESN reservoirs.
Local covariate information included indicators if the county bordered the ocean, was directly east (or west) of the Connecticut river, and if the county was not on the river, similar to \cite{hooten2010statistical}, as shown in Figure \ref{fig:rabiesCovNA}.
The spread is shown in Figure \ref{fig:CTspread} along with the last step forecast. Figure \ref{fig:CThpd} shows the full details of the last step forecast, with the lower and upper values of the 95\% HPD shown. 

Inference can be performed on the importance of the local covariates
The coefficients for the local covariates can be interpreted, but it should be noted that the addition of the reservoirs does mean that direct comparison to the results of \cite{hooten2010statistical} is not possible because of possible confounding between the reservoirs and the covariates.
The BESN-plus methodology consists of fitting and weighing multiple models. 
Computation of the HPD interval for a local covariate can be done by looking at the posterior draws for each model and weighing them by the learned model weights.
From these weighted draws from the posterior HPD intervals can be computed.
That said, the BESN-plus model shows that the coefficient for the indicator that the county was not bordering the river, the third panel of Figure \ref{fig:rabiesCovNA} was significant at the $95\%$ level and also negative, which can be interpreted as per the model specification. That is, if a county is bordering the river, the probability of that county having a presence of rabies is decreased - or in other words, the rate of rabies spread decreases in the presence of a river. This is similar to the results shown in \cite{hooten2010statistical}.

\section{Conclusion}
\label{sec:Conclusion}
Statistical estimation of stochastic CA spatio-temporal models can be computationally challenging \citep[e.g.,][]{banks2021statistical}. In the context of binary or categorical data, one can borrow from the GLMM-based spatio-temporal modeling literature to learn transition probabilities if the transformed mean response is reasonably modeled as linear and one has a good understanding of the necessary local transition rules (covariates).  However, with real-world spatio-temporal data, we seldom have a complete understanding of these local rules.  To compensate for such lack of knowledge, one can consider a latent Gaussian dynamic model \cite{grieshop2023datadriven}. Yet, many spatio-temporal processes evolve nonlinearly and latent linear dynamic models are not sufficient to capture the evolution.  Here, we consider a novel approach where we utilize reservoir-computing based ESNs as sources for the latent dynamics, in conjunction with the local covariates.

This BESN-plus method builds upon traditional ESN methods, which are typically applied to continuous output, or inappropriately assume categorical responses are continuous. ESN methods also do not have a natural way to accommodate prediction uncertainty. Our approach presents a proper handling of categorical (binary) data in an ESN framework, allowing the response to be properly modeled by a binary distribution as opposed to a simplified linear model. Embedding this within a Bayesian framework allows for uncertainty in the predictions, where previous methods utilize dropout or simple ensemble approaches to quantify uncertainty.  We also present an ensemble model averaging approach that mitigates the potential lack of expressiveness of any given reservoir from an ESN. That is,
weighting multiple reservoirs is used so that the random nature of the ESN procedure does not lead to poor performance if an unlucky reservoir is constructed. 
This use of reservoir ensembles also also allows for sampling different values of the tuning parameters, for which we provide an efficient way to generate informative prior distributions.
This method is parallelizable - depending on the amount of computational resources available more reservoirs can be constructed as each can be done independently from the others and then weighted once.

We demonstrate that our method can model the spread of a binary spatio-temporal process even when the local covariates are not specified properly. We also demonstrate that, although the methodology is most appropriate for prediction, it did suggest particular environmental covariates were important for the spread of raccoon rabies as has been demonstrated in the literature.

Importantly, BESN-plus method allows for the use of both  local covariate information and ESN dynamics. 
Our method can be extended to other data distributions such as multinomial, count, or continous data. Finally,
utilizing the reservoirs as merely a source of non-linear transformation of inputs (or stochastically generated basis functions) allows them to be used in virtually any traditional Bayesian method that can incorporate covariates with model selection \citet[e.g.,][]{Gelman_Carlin_Stern_Dunson_Vehtari_Rubin_2013}.

\newpage
\appendix

\section{Stan code} \label{ap:stan}
The following gives an example of the Stan code used for the implementation of the BESN-plus method.
The code requires that the output of the ESN is available. In the code this is denoted by $x$ and the associated output weight matrix is given the regularized horseshoe prior \citep{piironen2017sparsity}.
Additional spatio-temporal covariates are denoted by the variable $preds$ and correspond to the variables whose coefficients are assigned a normal prior (as is also true for the intercept parameters).
{\small
\begin{lstlisting}
data {
  int<lower=0> N; // number of spatial locations
  int<lower=0> H; // number of hidden units
  int<lower=0> R; // number of time points
  int<lower=0,upper=1> y[N,R]; // response variable
  matrix[H,R] x; // output from ESN
  matrix[N,R] preds; // additional covariates
  real<lower=0> scale_icept;
  real<lower=0> scale_global;
  real<lower=0> scale_b3;
  real<lower=1> nu_global;
  real<lower=1> nu_local;
  real<lower=0> slab_scale;   
  real<lower=0> slab_df;
}

parameters {
  real alpha;
  real b3;
  matrix[N, H] z;                
  real <lower=0> tau;         
  matrix<lower=0>[N, H] lambda; 
  real<lower=0> caux;
}

transformed parameters {
  real c = slab_scale * sqrt(caux);
  matrix[N, H] beta;
  matrix[N,R] theta;
  for(n in 1:N){
    for(h in 1:H){
      beta[n,h] = z[n,h] * sqrt(c^2*lambda[n,h]^2 / (c^2 + tau^2*lambda[n,h]^2));
    }
  }
  theta = beta * x;
}

model {
  // half-t priors for lambdas and tau, and inverse-gamma for c^2
  to_vector(z) ~ std_normal();
  to_vector(lambda) ~ student_t(nu_local, 0, 1);
  tau ~ student_t(nu_global, 0, scale_global*2);
  caux ~ inv_gamma(0.5*slab_df, 0.5*slab_df);
  alpha ~ normal(0, scale_icept);
  b3 ~ normal(0, scale_b3);
  for(n in 1:N){
    for(r in 1:R){
      target += bernoulli_logit_lpmf(y[n,r] | theta[n,r] + alpha + preds[n,r]*b3);
   }
 }
}
\end{lstlisting}
}

\newpage
\bibliographystyle{apalike}
\bibliography{references_clean}

\end{document}